
============================================================================
\magnification=1200
\hyphenpenalty=2000
\tolerance=10000
\hsize 14.5truecm
\hoffset 1.truecm
\openup 5pt
\baselineskip=18truept
\font\titl=cmbx12
\def\der#1#2{{\partial#1\over\partial#2}}


\def\rg{R_g}

\def\kes{\kappa_{es}}

\def\ref{\par\noindent\hangindent 20pt}
\def\refig{\par\noindent\hangindent 15pt}
\def\mincir{\raise -2.truept\hbox{\rlap{\hbox{$\sim$}}\raise5.truept
\hbox{$<$}\ }}
\def\magcir{\raise -4.truept\hbox{\rlap{\hbox{$\sim$}}\raise5.truept
\hbox{$>$}\ }}
\def\rho{\varrho}
\def\Mdot{\dot M}
\def\MS{M_*}

\null
\vskip 1.2truecm

\centerline{\titl DYNAMICALLY COMPTONIZED SPECTRA FROM}
\smallskip
\centerline{\titl NEAR CRITICAL ACCRETION ONTO NEUTRON STARS}
\vskip 1.5truecm
\centerline{Luca Zampieri}
\medskip
\centerline{International School for Advanced Studies, Trieste}
\centerline{Via Beirut 2--4, 34014 Miramare--Trieste, Italy}
\vskip 0.3truecm
\centerline{Roberto Turolla}
\medskip
\centerline{Department of Physics, University of Padova}
\centerline{Via Marzolo 8, 35131 Padova, Italy}
\vskip 0.3truecm
\centerline{and}
\vskip 0.3truecm
\centerline{Aldo Treves}
\medskip
\centerline{International School for Advanced Studies, Trieste}
\centerline{Via Beirut 2--4, 34014 Miramare--Trieste, Italy}
\bigskip\bigskip\bigskip\bigskip\bigskip
\centerline{Accepted for publication in the Astrophysical Journal}
\bigskip\bigskip\bigskip\bigskip\bigskip
\centerline{Ref. SISSA 96/93/A (July 93)}
\vfill\eject

\beginsection ABSTRACT

We investigate the effects of dynamical Comptonization on the emergent
radiation spectrum produced by near critical accretion onto a neutron star.
The flow dynamics and the transfer of radiation are self--consistently solved
in the case of a spherically symmetric, ``cold'', pure scattering flow,
including general relativity. A sequence of models, each characterized by the
value of the total observed luminosity, was obtained assuming that the
spectrum at the star surface is black body in shape. It is found
that when the luminosity
approaches the Eddington limit dynamical effects become important shifting
the spectrum to the blue and producing a power--law, high--energy tail.
The relevance of these results in connection with the observed spectral
properties of LMXBs, and of Cyg X--2 in particular, are discussed.
\bigskip\bigskip
\noindent
{\it Subject headings:\/} accretion \ -- \  radiative transfer \ -- \  stars:
individual\hfill\break
\phantom{\it Subject headings:\/} (Cyg X--2) \ -- \ stars: neutron \ -- \
X--rays: binaries
\vfill\eject

\beginsection I. INTRODUCTION

Most of the bright X--rays sources in the Galaxy are close binary systems
containing an accreting neutron star. In several cases the neutron star
manifests through a regular periodicity (X--ray pulsators), which is
related to the high magnetic field ($B\sim 10^{12} \ \rm {G}$) endowed by the
rotating neutron star.
However in the great majority of the so called Low Mass
X--ray Binaries (LMXBs) no regular modulation is present and the
neutron star magnetic field may be orders of magnitudes below that of
X--ray pulsators. In this case the dynamics of accretion onto the neutron star
may be practically unaffected by the magnetic field.

Observationally the spectrum of LMXBs is rather complex
and is  characterized by a drop in the
flux above $6\div 10$ keV, which  contrasts with the substantially harder
emission of X--ray pulsators; in some sources, however,
a steep power--law tail above
20 keV is observed (see e.g. Maurer {\it et al.\/} 1982, Matt {\it et
al.\/} 1990 and references therein). Fitting the observed spectra
requires the superposition of a number of simple laws, like a thin
bremsstrahlung plus a black body, or a bremsstrahlung plus a power law,
or a multitemperature black body disk plus a boundary layer black body,
or a Comptonized spectrum. Since LMXBs are variable, their emission can
be characterized by X--ray intensity--color and color--color
relations which have been studied in detail also in connection with the
so called Quasi Periodic Oscillations observed in a number of these
sources (see e.g. van der Klis 1989).
Basically one finds states in which the X--ray hardness ratio increases with
the
X--ray intensity, and states where it is substantially constant,
although the actual behaviour is much more complicated. The total luminosity
is, in some cases, a large  fraction of the Eddington limit for a
solar mass.

While a good fit to the spectral data obviously requires the superposition
of various physical regions, each  emitting a rather complex
spectrum, it  seems of interest to construct simple models, physically
well defined, which can shed light on some aspects of the formation of the
spectrum and be part of the overall picture.

In this spirit we consider here a spherically accreting, unmagnetized
neutron star which radiates close to the Eddington limit.
The hydrodynamics corresponding to such a picture was studied by
Maraschi, Reina \& Treves (1974, 1978), who found the noticeable result that
for high luminosities the infall velocity, approaching the star, reaches a
maximum and then starts to decrease,
owing to the pressure exerted by radiation; a settling regime is
established above the neutron star surface. This problem was
re--examined by Miller (1990) and Park \& Miller (1991), who included
also general relativity and obtained very similar results.

Radiation interacts with the inflowing gas and, even in the absence of
emission processes, the emergent spectrum
can be modified by dynamical Comptonization due to bulk motion
(Blanford and Payne 1981 a,b; Payne and Blanford 1981; Colpi 1988).
The effects of dynamical Comptonization  in accretion onto neutron stars were
recently investigated by Mastichiadis  and Kylafis (1992).
Their results, however, were obtained using an assigned, power--law
velocity profile, which is a poor approximation when the luminosity is
close to the Eddington limit,
and they are valid only in the diffusion regime.

In this paper we present self--consistent solutions for the transfer
of radiation in accreting flows onto a neutron star.
General--relativistic radiative transfer is tackled using
Projected Symmetric Trace Free (PSTF) moment formalism (Thorne 1981;
Turolla \& Nobili 1988; Nobili, Turolla \& Zampieri 1993).
In order to explore the effects of dynamics and gravity, we treat just
the pure scattering case and assume that photons are generated
at the neutron star surface with a Planckian spectral distribution.

The plan of the paper is the following. In section II we discuss the equations
governing the gas dynamics and the transfer of radiation while the results of
numerical integration are presented in section III. Section IV is
devoted to a comparison of our models with the observed spectral properties
of LMXBs.

\beginsection II. THE MODEL

We assume that X--ray radiation observed in LMXBs is generated by the
conversion of gravitational potential energy, as matter is spherically accreted
onto a slowly--rotating, weakly--magnetized neutron
star. If the flow velocity vanishes at the neutron star surface, the
efficiency of the accretion process is given by the variation of
the specific gravitational energy $E_p$
$$\epsilon = {(E_p)}_\infty - {(E_p)}_* = 1 - \left(\sqrt{-g_{00}}\right)_* =
1 - \sqrt{1 - {{2G\MS}\over{c^2R_*}}}\sim {{GM_*}\over{c^2R_*}},\eqno(1)$$
where $\MS$ and $R_*$ denote the neutron star mass and radius, respectively,
and we used vacuum Schwarzschild solution to describe the gravitational field
outside the star. By introducing the adimensional radial coordinate $r =
R/\rg$, $\rg = 2G\MS/c^2$,
and the accretion rate $\Mdot$, the total luminosity observed at infinity is
$$L_\infty = \epsilon \Mdot c^2 = \left(1-\sqrt{1-{1\over{r_*}}}
\right)\Mdot c^2 .\eqno(2)$$
The complete analysis of a steady--state, spherically symmetric gas flow
onto a compact star is a complex task
since the appearance of shocks and/or of a boundary layer at the neutron
star surface should
be expected. However, some reasonable simplifying assumptions can be made
if one is not interested in treating
in detail the inner accretion layer where nearly all the energy is released.
First of all we note that $\epsilon\sim 0.1$ for $\MS\sim1.5M_\odot$ and
$R_*\sim 10$ km, so that
$L_\infty/L_E= l_\infty\sim 1$ if $\Mdot/\Mdot_E=\dot m\sim 1$, where
$L_E = 4\pi G\MS c/\kes$ is the Eddington luminosity and $\Mdot_E = L_E/c^2$ is
the critical accretion rate.
By comparing this value for $l_\infty$ with the maximal luminosity attainable
in
black hole accretion, $l_{BH}\mincir 0.01$ (see Nobili, Turolla \& Zampieri
1991), it follows that the infalling gas can radiate, at most, a few percent
of the total energy output before the impact with the surface of the star.
We can therefore safely neglect emission processes in the accreting gas and
treat the material as a pure scattering medium. In this hypothesis the
radiation spectrum observed at infinity is formed in the boundary layer
by processes which are not important to specify in detail since the overall
spectral properties are determined mainly by
scatterings as radiation propagates outward. The main goal of our
investigation is to study the effects of dynamics on radiative transfer and
therefore we consider only coherent (Thomson) scattering.
We make the further assumption that the plasma is ``cold'', in the sense
that its enthalpy is always much less than effective gravity. This is
equivalent to say that the the gas velocity is everywhere greater than the
sound
speed, $v > v_s$, and limits the validity of our approach to the supersonic
part of the flow.

For a recent derivation of the equations governing the dynamics of the matter
gas and the transfer of radiation in spherical, stationary accretion in a
Schwarzschild gravitational field we refer to
Nobili, Turolla \& Zampieri (1991). In the case of a ``cold'', pure scattering
plasma they reduce to the form

$$v^2{{(yv)^{'}}\over{yv}} = - {1\over{2y^2r}} + {{4 \pi \kes r\rg}\over
{yc^2}}
 H,\eqno(3)$$

$$2\kes \rg r^2\rho vy = \dot m,\eqno(4)$$

$$ \eqalignno{
 & H^{'} - v J^{'} + 2 H \left( 1 + {y^{'}\over y} \right)
   - v J \left[ f \left ( {{(yv)}^{'}\over {yv}} - 1 \right)+
{4\over 3} \left( {{(yv)}^{'}\over {yv}} + 2 \right)\right] = 0 & (5a)\cr
  &\left(f +{1\over 3}\right) J^{'} - v H^{'}
   + \left[ f^{'} + f \left(3 + {y^{'}\over y} \right)+
 {4\over 3} {y^{'}\over y}\right] J
   - 2vH \left[ {{(yv)}^{'}\over {yv}} + 1 \right]
    =\cr
 & = -{{\kes r\rg\rho H}\over y}\ . &(5b)\cr } $$
Here $v$ is in units of $c$, $y=\sqrt{(1-1/r)/(1-v^2)}$, a prime denotes
derivation with respect to $\ln r$ and $J$ and  $H$ are the radiation energy
density and flux, as measured in the local rest frame (LRF) of the fluid; the
comoving luminosity $L$
is related to $H$ by $L=16 {\pi}^2 r^2\rg^2cH$, where an extra $c$ factor
appears because all moments are in $\rm erg/cm^3$.
The variable Eddington factor, $f= K/J - 1/3$ is a given function of the
optical depth $\tau$ of the form
$$f(\tau ) = {2\over{3(1+\tau^2)}}\, .\eqno(6)$$

The solution of equations (3)--(5) provides the velocity and density
profiles of the matter gas and the radial evolution of the
frequency--integrated moments, once
boundary conditions are given. In  treating a pure scattering problem
one of the conditions must fix the value of either the radiation energy density
or of the flux at some radius. Computed models were obtained assigning the
value
of the luminosity far from the star. Since $\MS$ and $R_*$ are known, equation
(2) gives immediately the accretion rate, while the two remaining boundary
conditions are
$$J=H\qquad\qquad r=r_{out}\qquad\qquad\qquad ({\rm radial \
streaming})$$
and
$$v= \sqrt{{{1-l_\infty}\over r}}\qquad\qquad r=r_{out}\qquad\qquad\qquad (
\hbox{\rm
``modified'' free fall}).$$
It should be noted that $l_\infty$ is the only free parameter of the model.

The results of numerical integrations are shown in figures 1 and 2, for two
representative values of $l_\infty$, $l_\infty = 0.3$ and $l_\infty = 0.9$
respectively; in all models $\MS = 1.4 \ M_\odot$ and $R_* = 10$ km.
The settling regime is a common feature of all high luminosity solutions
and is clearly visible in figure 2b. The overall behaviour of our
models is close to that found by Maraschi {\it et al.\/} (1978), Miller
(1990) and Park \& Miller (1991) under quite similar assumptions.

The self--consistent velocity and density profiles obtained from the
simultaneous integration of the hydrodynamical and frequency--integrated
moment equations can
now be used to solve the full, frequency--dependent transfer problem. The
frequency--dependent version of equations (5) is (Thorne 1981;
Nobili, Turolla \& Zampieri 1993)

$$\eqalignno{&
\der{H_\nu}{\ln r} +
2H_\nu + {{y^{'}}\over{y}} \left(
H_\nu - \der{H_\nu}{\ln\nu} \right)
-v \left\{\der{J_\nu}{\ln r}
 +\left(\beta +2\right)J_\nu - \right. & \cr
&\left. \left[f\left( \beta - 1\right)
- {1\over 3} \left(\beta +2\right)\right] \der{J_\nu}{\ln\nu}\right\}
= 0 & (7a)\cr }
$$

$$\eqalignno {&
\left( f + {1\over 3}\right)\der{J_\nu}{\ln r} + f\left( 3 + {{f^{'}}\over f}
\right)J_\nu
+ {{y^{'}}\over{y}}\left[ J_\nu - \left( f +{1\over 3}\right)
\der{J_\nu}{\ln\nu}\right]- &\cr
& v\left[\der{H_\nu}{\ln r} + {1\over 5}\left(7\beta +8\right)
H_\nu - {1\over 5}\left(3\beta + 2 \right)\der{H_\nu}{\ln\nu}\right]
- g (\beta - 1) \left( H_\nu + \der{H_\nu}{\ln\nu} \right) = \cr
&  -{{\kes\rho r\rg H_\nu}\over{y}}\, ,
&(7b)\cr}$$
where $\beta = (yv)^{'}/(yv)$ and the index $\nu$ denotes
frequency--dependent quantities. In writing equations (7) we have assumed
that the Eddington factors $f$ and $g = N/H - 3/5$
are independent of frequency;
actually $f$ is given again by equation (6) and $g = 3f/5$.

Equations (7) have to be solved as a two points
boundary value problem in space and an intial value problem in frequency. In
our particular case, we have assigned the flux spectral distribution at the
star surface
$$H_\nu(r_*) = \bar H_\nu$$
with the additional constraint that
$$\int_0^\infty \bar H_\nu\, d\nu = H(r_*)\, ,\eqno(8)$$
where $H(r_*)$ is the frequency--integrated flux at $r_*$, as given by the
solution of equations (3)--(5); the remaining radial condition is again
$$J_\nu = H_\nu\qquad\qquad r=r_{out}\qquad\qquad\qquad ({\rm radial \
streaming}).$$

A detailed discussion of the form of the input spectrum, together with
the choice of the frequency conditions, is postponed to the next section.
Here we want to show that dynamical effects on radiative transfer are going
to be  relevant in our models. As discussed by Payne \& Blandford (1981)
and Nobili, Turolla \& Zampieri (1993), bulk motion Comptonization is expected
to become important in regions where $\tau >1$ and $\tau v\sim 1$. By making
use of equations (2), (4), neglecting relativistic corrections and assuming
$\tau\sim \kes\rho r\rg$, we have that
$$(\tau v)_*\sim {{\dot m}\over{2r_*}}\sim l_\infty\, ,\eqno (9)$$
which shows that close to the star surface $\tau v$ is indeed not far from
unity for near critical accretion. Beside this effect, which is not
relativistic arising from first order terms in $v$, GR corrections are also
important, at least when velocity starts to deviate from free fall. In this
case, in fact, $y$, which is very close to unity and nearly constant in free
fall, drops below one and photons start to experience a gravitational
redshift.

\beginsection III. RESULTS

In the case of a weakly--magnetized neutron star accreting close to the
Eddington limit, as we are considering here, we
assume that photons are generated only at
the surface, through some convenient mechanism that could convert
the kinetic energy of the infalling gas into radiation with a given spectral
distribution. The simplest
choice that can be made is that the input spectrum is  Planckian in shape
$$ H_x (r_{*}) = A \ {{{x}^3}\over {\exp{x} - 1}} \, ,\eqno(10)$$
$x=h\nu/kT_*$, where the temperature $T_{*} = T(r_{*})$ is fixed by the
condition
$$ T_{*} = {\left( {{L_{\infty}}\over {4 \pi R_{*}^2 \sigma}} \right)}^{1/4}.
\eqno(11)$$
This expression for $T_*$ is justified if the effective optical depth in the
flow is always smaller than unity, as indeed should be the case in all our
models. The constant $A$ appearing in equation (10) is determined
by normalization (8).

The integration of equations (7) was performed using an original numerical
code based on a relaxation method (for a detailed discussion
see Nobili, {\it et al.\/} 1993). This transfer code was designed to deal with
fairly general situations in which dynamics and/or gravity become important
and was tested solving a number of simple problems for which analytical
solutions were available (Nobili, {\it et al.\/} 1993).
As a further check on the accuracy of the method
in the present case, we compared the integral of the frequency dependent flux
with the integrated luminosity at each radius; the largest relative error is
some percent. A 30 frequency bins $\times$ 60 radial zones grid
with $0.4 < \log r < 5$ and $- 0.5 < \log x < 1.2$ was used.

Two different sets of frequency boundary conditions
have been used in the calculations, according to the value of $l_\infty$ which
characterizes the model. For models with luminosity $l_{\infty}\mincir 0.4$ the
spectrum drifts toward low frequencies because gravitational redshift
dominates over dynamical Comptonization (see the following discussion), so
that both conditions must be specified at the largest frequency mesh point
$x_{max}$
$$ \eqalignno{
& \der{\ln J_{x}}{\ln x} = \der{\ln H_{x}}{\ln x} \ \ \ \ \
\ \ \ \ \ x = x_{max}  &(12a)\cr
& \der{\ln J_{x}}{\ln x} = 3 - x \ \ \ \ \
\ \ \ \ \ x = x_{max} \ \ \ \ \ \ ({\rm Wien \ law}). &(12b)\cr}$$
The situation is reversed for models with luminosity $l_{\infty}\magcir 0.4$
and (12) are replaced by
$$ \eqalignno{
& \der{\ln J_{x}}{\ln x} = \der{\ln H_{x}}{\ln x} \ \ \ \ \
\ \ \ \ \ x = x_{min}  &(13a)\cr
& \der{\ln J_x}{\ln x} = 2 \ \ \ \ \
\ \ \ \ \ x = x_{min} \ \ \ \ \ \ (\hbox{\rm Rayleigh--Jeans law}).
&(13b)\cr
}$$
Conditions (12b) and (13b) express the fact that $J_x$ must remain black
body in shape at high (low) frequencies while (12a) and (13a) derive from the
request that $H_x\propto J_x$ in a pure scattering medium. A
discussion on how frequency conditions should be placed in solving
the transfer problem in moving media can be found in Mihalas, Kusnaz
\& Hummer (1976), Nobili {\it et al.\/} (1993), Turolla, Nobili \& Zampieri
(1993).
Numerical integrations show that in a small luminosity range around $l_\infty
= 0.4$ both sets of boundary conditions work satisfactorily.

Some representative spectra are shown in figures 3 and 4 and results are
summarized in table 1, where the ``soft'' and ``hard'' colors of the emergent
spectrum are reported together with count rate in the 1--17 keV energy band;
the last column gives the value of the photon spectral index if a power--law
tail forms at high energies.
The number flux was computed assuming a distance of 8 kpc for the source.

As can be clearly seen from the figures, the behaviour of the emergent spectrum
changes significantly around $l_\infty = 0.4$. For $l_\infty\mincir 0.4$, in
fact, the flow becomes optically thin to scattering and,
although $(\tau v)_*\sim 0.3$, bulk motion Comptonization has little effect,
because the probability for a photon to scatter before escaping to infinity
becomes very low. The fact that $(v\tau )_*$ is not far from unity
means that the typical fractional energy change per scattering
is still large but electron--photon collisions are so few that the total energy
exchange is negligible.
Consequently the spectrum at infinity remains nearly Planckian but is rigidly
shifted to the red (see figure 3) by the effect of gravity.
As $l_\infty$ is increased beyond 0.4, models
start to develop a thick core, $(\tau v)_*$ remains fairly constant because
$v_*$ decreases and $\tau$ increases, but now dynamical Comptonization is
important owing to the larger optical depth near the stellar surface. The
emergent spectrum is
systematically shifted to the blue and a power--law, high--energy tail forms.
The dynamical blueshift roughly compensates the gravitational redshift just
at $l_\infty\sim 0.4$. A typical dynamically Comptonized spectrum is shown
in figure 4 for $l_\infty = 0.9$. The spectral shape is essentially Planckian
up to energies $\sim 20$ keV and then becomes a power law with spectral index
$\alpha\simeq 3.4$. We wish to stress that low  and high luminosity models
have different spectral properties because the former are everywhere
optically thin (or at most marginally thick) while the latter do have an
optically thick inner region. The fact that $(v\tau)_*$ is more or less the
same for all the solutions does not contradict the previous statement. In fact,
from the analysis of Payne \& Blandford it follows that $v\tau$ is the
important parameter for dynamical Comptonization, but it should be taken into
account that their results were obtained in the diffusion limit, that is to say
in the hypothesis of very large scattering depth. For optically thin flows
Payne \& Blandford approach does not apply at all and one has to expect the
emergent spectrum to show different features. On the other hand, when
$l_\infty\magcir 0.4$,
spectra originating from near critical accretion onto neutron stars
closely resemble those obtained by Payne \& Blandford (PB)
for spherical accretion onto black holes. PB have shown,
in fact, that a monochromatic line at $\nu = \nu_0$,
injected at a basis of a converging fluid flow, experiences both a drift
and a broadening toward frequencies higher than $\nu_0$, as photons propagate
outward. This effect is due to repeated Thomson scatterings by the moving
electrons and is intrinsically similar to first order Fermi acceleration of
cosmic rays. If the flow is characterized by a constant velocity gradient
$\beta = v^{'}/v$, the emergent number flux $N_\nu = H_\nu/h\nu$ shows a
power--law, high--energy tail with spectral index
$$\alpha = -{{\partial{\ln N_\nu}}\over{\partial{\ln \nu}}} =
{{5+\beta}\over {2+\beta}}\, .$$
Recently Mastichiadis \& Kylafis (1992, MK) extended PB's
analysis to an accreting flow onto a neutron star.
Essentially they performed again
PB's calculations, but replaced the original boundary condition
with the
requirement that the flux vanishes at the star surface and proved that now
$\alpha = 1$, irrespective of the value of $\beta$. While the overall
spectral evolution is quite similar in both cases, some important points,
like the value of the final power--law index, seem to depend on the
choice of the condition for the radiation field at the inner boundary.
This particular aspect deserves a further comment in order to understand how
our numerical solutions relate to these analytical results. Both PB and MK
solved a Fokker--Planck equation for the photon occupation number, $n(t,\nu)$,
$t = 3\tau v\propto R^{-1}$,
looking for separable solutions of the form $n = f(t)t^{3-\beta}\nu^{-p}$.
The resulting ordinary differential equation
for $f(t)$ is a Kummer equation with coefficients depending both on
$\beta$ and $p$. To select the unique well--behaved, physical solution,
boundary conditions must be provided and here the two approaches differ
because the geometry changes. While PB considered a converging flow which
extends up to $R=0$  and could require regularity for
$t\to\infty$, MK were forced to use a different condition (vanishing flux) at
$R=R_*$ where their integration domain terminates with a finite value of $t$.
The two conditions at the basis of the flow produce two distinct
sets of eigenvalues $p_n$ and, since $\alpha = p_0$ if the effect is saturated,
this explains the different behaviour at large frequencies, although the input
spectrum is a monochromatic line in both cases.
Actually, in the model we are considering,
the integration domain is finite, as in the MK case,
so that PB's analysis certainly does not apply; our approach, however differs
also from that of MK, because
in fixing the spectrum at $R_*$, we have assigned only the photon input,
without specifying any extra constraint.
In this situation we do not expect to select any countable set of eigenvalues.
The emergent spectrum shows, nevertheless, a well defined, power--law tail,
whose index seems
to be related to the physical conditions in the region where dynamical
effects become important.

In concluding we note that, although present results refer to the radial
evolution of a Planckian, the
global properties of the model (drift to high frequencies and formation of
a power law tail) are largely independent of the form chosen for the input
spectrum.

\beginsection IV. DISCUSSION

Here we consider the relevance of our models in connection
with the observed spectral properties of accreting neutron stars, such as
those  believed to power LMXBs.
In comparing our synthetic spectra with actual observations one should keep
in mind the drastic simplifications
introduced in the model itself. We have considered a unique component,
a dynamically Comptonized black body, and in treating radiative transfer we
have neglected all emission--absorption process so that, for instance, line
features do not appear.

In order to be specific we shall deal with the spectral observations of
Cyg X--2, which is one of the prototypes of the class of LMXBs. We refer in
particular to EXOSAT observations of 1983, which covered the energy band
0.5--20 keV (see Chiappetti {\it et al.\/} 1990) and  focus our attention
on the observations of September 13--22. Assuming a  distance of 8 kpc
(Cowley {\it et
al.\/} 1979), the X--ray luminosity is $1.7\times 10^{38}$ erg/s, which clearly
situates the source close to the Eddington limit for $M_*\sim M_\odot$.
Note that Cyg X--2 is one of the few LMXBs with such a high luminosity.
A good fit to the data is formally obtained by the
superposition of a bremsstrahlung at 4.4 keV and a black body at 1.2
keV. We consider also the high energy observations of Maurer {\it et al.\/}
(1982), obtained during a balloon flight in
May 1976, which detected the source in the 18--60 keV range, with the
low energy point at a level comparable to that of the EXOSAT observations.
The high energy count rate is  well fitted by a power law of spectral index
$2.8\pm 0.7$. As mentioned above,
a useful tool for studying the X--ray spectral properties
of LMXBs are the intensity--color and color--color diagrams.
Cyg X--2 was, in fact, the first source for which
the constancy of the hardness ratio with  respect to
the intensity was discovered (Branduardi {\it et al.\/} 1980).
This rather puzzling behaviour, together with the characteristic Z--shaped
track on the X--ray color--color plot, is shared by many other LMXBs; more
details can be found e.g. in Hasinger (1987) and references
therein.
In figure 5 we have plotted the observations of
Chiappetti {\it et al.\/} and Maurer {\it  et al.\/} (errors below 20 keV
are  $\mincir 10\%$) together with the computed spectrum for the model with
$l_\infty=0.9$,
while a typical intensity--color plot for Cyg X--2, taken again from
Chiappetti {\it et al.\/}, is shown in figure 6. The color is defined as
the ratio of the number counts in the 6--17 and 3--6 keV bands while the
intensity refers to the count rate in the 1--17 keV range.

A comparison of the synthetic spectrum
with the observed one (see figure 5) clearly indicates
that our model can not describe the number counts in the entire
energy range because it exhibits a definite photon deficit at low frequencies.
However, the flux energy distribution
above  10 keV is a steep power law and it is well fitted by
the model; the computed spectral index, $\alpha\simeq 3.4$, is, in fact,
consistent within the large uncertainties with the observed one.

The intensity--color diagram for our solutions is plotted in figure 7.
As can be seen, for luminosities $l_\infty < 0.4$ the count rate in the 1--17
keV band increases, while, for $l_\infty > 0.4$, it starts
to (slowly) decrease, because the spectrum becomes harder and
the total number of photons in the interval 1--17 keV becomes lower. This
contrasts with the nearly linear dependence of the hardening ratio on
the total luminosity (see table 1).
The turning point in the diagram corresponds to the model with
$l_{\infty} \simeq 0.4$ which is the value of the luminosity at which
the count rate is maximum. We note that the existence of an anticorrelation
branch in the intensity--color diagram for our solutions is entirely due
to dynamical effects.
In the absence of bulk motion
Comptonization, in fact, the spectrum remains Planckian and drifts
to higher frequencies for increasing $l_\infty$ just because
$T_*\propto l_\infty^{1/4}$; the gravitational redshift does not
enter in these considerations being the same for all the models.
It can be easily checked that a Planckian spectrum produces
always a positive intensity--hardness correlation when temperature is
increased.
The turning point appears only when dynamical effects start to be important;
in this case the mean photon frequency
$\bar{\nu} \propto f_b$, where $f_b > 1$ is the dynamical blueshift
factor, and, as luminosity approaches $L_E$, $f_b$ can become
considerably large causing a decrease in the count rate,
while colors continue to increase because the spectrum becomes harder.

If we explain the X--ray variability of LMXBs in terms of a time
varying accretion rate,
the turning point present in the intensity--color diagrams of some LMXBs,
see e.g. van der Klis (1990), bears some resemblance with the
behaviour of our solutions.
In this respect one is tempted to interpret observations 1,2 and 4,5 of
figure 6 in the same terms, as corresponding to the models with
luminosities in the range 0.4--0.5 of figure 7.

In conclusion, although the simple model considered here cannot describe
the overall X--ray emission of Cyg X--2, since the computed spectrum
shows an evident deficit of photons at low frequencies, we find that dynamical
effects due to bulk gas motion (drift to high frequencies and formation of a
power law tail) may be an important ingredient to explain
some of the observed spectral properties of neutron stars accreting close
to the Eddington limit. The construction of a more complete model
which takes into account
also for emission processes in the infalling gas in addition to photons
produced at the star surface is presently under consideration.

\vfill\eject

\beginsection REFERENCES

\ref{Blandford, R.D., \& Payne, D.G. 1981a, MNRAS, 194, 1033.}
\ref{Blandford, R.D., \& Payne, D.G. 1981b, MNRAS, 194, 1041.}
\ref{Branduardi, G., Kylafis, N.D., Lamb, D.Q., \& Mason, K.O. 1980, ApJ,
235, L153.}
\ref{Chiappetti, L., Treves, A., Branduardi--Raymont, G., Ciapi, A.L.,
Ercan, E.N., Freeman, P.E., Kahn, S.M., Maraschi, L., Paerels, F.B.,
\& Tanzi, E.G. 1990, ApJ, 361, 596.}
\ref{Colpi, M. 1988, ApJ, 326, 223.}
\ref{Cowley, A.P., Crampton, D., \& Hutchings, J.B. 1979, ApJ, 231, 539.}
\ref{Hasinger, G. 1987, in IAU Symposium 125, The Origin and evolution of
Neutron Stars, (Dordrecht:Reidel), p. 333}
\ref{Maraschi, L., Reina, C., \& Treves, A. 1974, A\&A, 35, 389.}
\ref{Maraschi, L., Reina, C., \& Treves, A. 1978, A\&A, 66, 99.}
\ref{Mastichiadis, A., \& Kylafis, N.D. 1992, ApJ, 384, 136.}
\ref{Matt, G., Costa, E., Dal Fiume, D., Dusi, W., Frontera, F.,
\& Morelli, E. 1990, ApJ, 355, 468.}
\ref{Maurer, G.S., Johnson Neil, W., Kurfess, J.D., \& Strickman, M.S. 1982,
ApJ, 254, 271.}
\ref{Mihalas, D., Kusnaz, P.B., \& Hummer, D.G. 1976, ApJ, 206, 515.}
\ref{Miller, G.S. 1990, ApJ, 356, 572.}
\ref{Nobili, L., Turolla, R., \& Zampieri, L. 1991, ApJ, 383, 250.}
\ref{Nobili, L., Turolla, R., \& Zampieri, L. 1993, ApJ, 404, 686.}
\ref{Park, M.--G., \& Miller, G.S., 1991, ApJ, 371, 708.}
\ref{Payne, D.G., \& Blandford, R.D. 1981, MNRAS, 196, 781.}
\ref{Thorne, K.S. 1981, MNRAS, 194, 439.}
\ref{Turolla, R., \&  Nobili, L. 1988, MNRAS, 235, 1273.}
\ref{Turolla, R., Nobili, L., \& Zampieri, L. 1993, preprint.}
\ref{van der Klis, M. 1989, ARA\&A, 27, 517.}
\ref{van der Klis, M. 1990, in Neutron Stars: Theory and Observations,
(Dordrecht: Kluwer Academic Publishers), p. 319.}

\vfill\eject
%

\null
\vskip 1.5truecm
\centerline{Table 1}\medskip
\centerline{Characteristic Parameters for Selected Models}\bigskip\bigskip
$$\vbox{\tabskip=1em plus2em minus.5em
\halign to\hsize{#\hfil &\hfil # \hfil & \hfil # \hfil & \hfil # \hfil &
 \hfil # \hfil  & \hfil # \hfil & \hfil # \hfil &
\hfil # \hfil & \hfil # \hfil &\hfil # \hfil &\hfil # \hfil\cr
& $l_\infty $ & $\tau_*^{\rm a}$ & $(\tau v)_*$ & ${{\hbox{\rm (3--6)}}\over
{\hbox{\rm (1--3)}}}^{\rm b}$ & ${{\hbox{\rm (6--17)}}\over
{\hbox{\rm (3--6)}}}^{\rm c}$ & ${\rm  count \ rate}^{\rm d}$ &
$\alpha^{\rm e}$ & & & \cr
\noalign{\smallskip}
%
%
\noalign{\bigskip\medskip}
& 0.1 & 0.3  & 0.17 & 0.17  & 0.02 & 0.40 &- & & &   \cr
& 0.2 & 0.7  & 0.31 & 0.31  & 0.06 & 0.73 &- & & &   \cr
& 0.3 & 1.3  & 0.37 & 0.40  & 0.09 & 0.96 &- & & &   \cr
& 0.4 & 2.2  & 0.38 & 0.68  & 0.20 & 1.00 &- & & &   \cr
& 0.5 & 4.0  & 0.38 & 1.06  & 0.49 & 0.87 &- & & &   \cr
& 0.6 & 7.1  & 0.39 & 1.28  & 0.70 & 0.86 &- & & &   \cr
& 0.7 & 14   & 0.38 & 1.44  & 0.98 & 0.82 &- & & &   \cr
& 0.8 & 33   & 0.34 & 1.63  & 1.23 & 0.79 & 3.3 & & &   \cr
& 0.9 & $1.8\times 10^2$  & 0.30 & 1.79  & 1.53 & 0.75 & 3.4 & & &   \cr
& 0.95& $1.2\times 10^3$  & 0.28 & 1.87  & 1.69 & 0.72 & 3.5 & & &   \cr
 }}$$

\bigskip
\parindent 0.truept
$^{\rm a}$ electron scattering optical depth at the star surface

$^{\rm b}$ ``soft'' color: $N_{3-6}/N_{1-3}$, where $N$ is the
photon count rate in the specified energy range

$^{\rm c}$ ``hard'' color: $N_{6-17}/N_{3-6}$

$^{\rm d}$ count rate: $N_{1-17}$ (arbitrary units)

$^{\rm e}$ photon spectral index: $ - \partial\ln N_\nu/\partial\ln\nu$,
calculated above 20 keV

%
%
%
\vfill\eject

\beginsection FIGURE CAPTIONS

\refig{Figure 1a.\quad Velocity and density profiles versus radius
for the model with $l_\infty = 0.3$. Scales are logarithmic and density is
in g/cm$^3$.}
\medskip
\refig{Figure 1b.\quad Stationary, $l_s$, and comoving, $l$, luminosities
for the model with $l_\infty = 0.3$.}
\medskip
\refig{Figure 2a.\quad Same as in figure 1a for the model with $l_\infty =
0.9$}
\medskip
\refig{Figure 2b.\quad Same as in figure 1b for the model with $l_\infty =
0.9$.}
\medskip
\refig{Figure 3.\quad Number flux spectral distribution
for the model with $l_\infty = 0.3$. The source is assumed to be
at a distance of 8 Kpc.}
\medskip
\refig{Figure 4.\quad  Same as in figure 3 for the model with $l_\infty =
0.9$.}
\medskip
\refig{Figure 5.\quad The spectrum of Cyg X--2 in the 1--60 keV range.
The dotted line is the best fit to the data of Chiappetti {\it et al.\/},
crosses refer to the observations of Maurer {\it et al.\/} and the continuous
line is the model spectrum for $l_\infty = 0.9$.}
\medskip
\refig{Figure 6.\quad Hardness ratio versus count rate for
Cyg X--2, as from Chiappetti {\it et al.\/}; diamonds
mark in the plot the areas covered by five different groups of observations.}
\medskip
\refig{Figure 7.\quad  Same as in figure 6 for our models; each point is
labeled by the value of $l_\infty$ (see also table 1).}

\vfill\eject

\bye